\documentclass
[aps,prc,amsmath,amssymb,floatfix,nofootinbib]
{revtex4}

\usepackage[dvips]{graphicx}         %
\usepackage{epstopdf}
\usepackage{bm}                      
\usepackage{mathptmx}                
\usepackage{dcolumn}                 
\usepackage{multirow}                %
\usepackage[toc,page,title,titletoc,header]{appendix}

\def\bea{\begin{eqnarray}} \def\eea{\end{eqnarray}}
\def\beq{\begin{equation}} \def\eeq{\end{equation}}
\def\bal#1\eal{\begin{align}#1\end{align}}
\def\bse#1\ese{\begin{subequations}#1\end{subequations}}
 
\def\kv{\bm{k}}

\begin{document}

\title{Medium polarization effects in $^3SD_1$ spin-triplet pairing}

\author{
Wenmei Guo,$^{a,b}$ U. Lombardo,$^{b}$ \footnote{Corresponding
author lombardo@lns.infn.it at: Laboratori Nazionali del Sud
(INFN), Via S. Sofia 62, 95123 Catania, Italy, phone: +39\ 095\
542\ 277, fax: +39\ 095\ 71\ 41\ 815} P. Schuck$^{c,d}$}
\affiliation{ $^{a}$ Institute of Theoretical Physics, Shanxi University,
030006 Taiyuan, China \\
$^{b}$ Laboratori Nazionali del Sud (INFN),
Via S. Sofia 62, 95123 Catania, Italy\\
$^{c}$ Institut de Physique Nucl$\acute{e}$aire,
Universit$\acute{e}$ Paris-Sud, F-91406 Orsay Cedex, France\\
$^{d}$LPMMC (UMR5493), Universit$\acute{e}$ Grenoble Alpes and
CNRS, 25, rue des Martyrs, B. P. 166, 38042 Grenoble, France}
\date{\today}

\begin{abstract}
Stimulated by the still puzzling competition between spin-singlet
and spin-triplet pairing in nuclei, the $^3SD_1$ neutron-proton
pairing is investigated in the framework of BCS theory of nuclear
matter. The medium polarization effects are included in the single
particle spectrum and also in the pairing interaction starting
from the $G$-matrix, calculated in the Brueckner-Hartree-Fock
approximation. The vertex corrections due to spin and isospin
collective excitations of the medium are determined from the
Bethe-Salpeter equation in the RPA limit, taking into account the
tensor correlations.  It is found that the self-energy corrections
confine the superfluid state to very low-density, while remarkably
quenching the magnitude of the energy gap, whereas the induced
interaction has an attractive effect. The interplay between
spin-singlet and spin-triplet pairing is discussed in nuclear
matter as well as in finite nuclei.
\end{abstract}



\maketitle

\section{Introduction}
For several decades the strong experimental evidence of the
spin-singlet pairing between like nucleons in nuclei has been
stimulating intense theoretical activity until the recent years
\cite{broglia}. 
On the contrary, there is not yet a clear evidence
for neutron-proton (np) spin-triplet pairing. 
This is the reason why this kind of pairing has received much less 
attention \cite{satula,poves,goodm99}. 
However, it is well known since long
that the $T=0$ np interaction could give  relevant pairing, 
being more attractive than the $T=1$ interaction \cite{lane}. 
In recent calculations on the competition between spin-singlet and
spin-triplet pairings in N=Z nuclei it has been argued that the
latter is hindered by the spin-orbit splitting
\cite{bertsch10,bertsch12,sagawa13}. 
However, in Ref.~\cite{bertsch10} it is pointed out that in very large N=Z
nuclei ($A>140$) spin-triplet pairing condensates are favored
because the spin-orbit force becomes vanishing small. 
The disappearance of the $S=1$, $T=0$ pairing  with asymmetry in nuclei has
been studied in, e.g., Ref.~\cite{schu}. 
In those calculations no dynamical effects on pair correlations are considered, 
whereas it is proved that particle-vibration coupling could yield a
significant contribution to the pairing gap magnitude in the
spin-singlet case \cite{barranco} and also in the neutron-proton
spin-triplet one, even if less significant \cite{litvinova}. 
On the other hand, in the vicinity of the proton drip in heavier
nuclei the spin-triplet pairing could potentially also become more
important. 
This may be revealed by further theoretical and
experimental investigations.

Studies of neutron-neutron (nn) and proton-proton (pp) pairing in
nuclear matter have also addressed the medium collective
excitations \cite{cao06,zhang}, which can enhance or quench the
the pairing correlations according to the nuclear environment
where the Cooper pairs are embedded. In the case spin-singlet nn
pairing in symmetric nuclear matter the medium-induced interaction
significantly enhances the gap, supporting calculations of energy
gaps in $^1S_0$ neutron-neutron (nn) or proton-proton (pp)
spin-singlet pairing in nuclei, where pair vibrations are
included \cite{barranco}.

In the case of spin-triplet np pairing  BCS calculations with bare
interaction in nuclear matter predict sizable energy gaps of the
order of 12 MeV, i.e. four times that of the spin-singlet
\cite{lombardo}. Even if significant re-scaling is expected from
the self-energy effects, the energy gap could be still large
enough by anti-screening due to the induced interaction
\cite{cao06}. Therefore, the predicted effect of the spin-orbit
energy splitting could be resized by the large spin-triplet pair
correlation energy.

In this paper we discuss the $^3SD_1$ spin-triplet np pairing in
symmetric nuclear matter, taking into account both self-energy
insertions to the quasi-particle spectrum and vertex corrections
to the bare interaction due to collective excitations of the
medium. The vertex corrections have been determined from the RPA
version of the Bethe-Salpeter (BS) equation in the Landau limit.
However the RPA does not consider the feedback of the effective
interaction on the collective modes, i.e. dressing the
polarization propagator with the full interaction in a
self-consistent procedure \cite{babu,indint}.

As leading term the Brueckner-Hartree-Fock (BHF) $G$-matrix is
adopted to prevent the divergences due to hard-core of the nuclear
force. The strong tensor force present in the bare interaction
deeply affects the $G$-matrix so that it cannot be neglected. This
entails that the tensor parameters must be included in the effective
interaction, solution of the BS equation, when expressed in terms
of Landau-Migdal parameters \cite{migdal}.
 The resulting energy gap will be compared with the
$^1S_0$ spin-singlet nn (or pp) gap and estimates, based on the
local density approximation (LDA) will be made for the gaps in
nuclei.
%
%
\section{Theoretical framework}
\subsection{Gap equation}
In this section, the formalism of the BCS theory of the $^3SD_1$
superfluid state of symmetric nuclear matter is set, including the
medium polarization effects \cite{lombardo,schulze}. The two
coupled gap equations ($L=0, 2$) are written as\beq \Delta^{ST}_L(k)
= -\frac{Z^2_F}{\pi}\int^{\infty}_{0} k'^2 dk' \sum_{L'}
\frac{V^{ST}_{LL'}(k,k')}{\sqrt{\varepsilon^2_{k}+\Delta(k')^2}}
\Delta^{ST}_{L'}(k'), \eeq where \beq \Delta(k)^2 =
\Delta^{ST}_0(k)^2 +\Delta^{ST}_2(k)^2, \eeq

The prefactor $Z_F$ is the quasi-particle strength which  takes
into account the depletion of the Fermi surface \cite{Zdick}. The
quasi-particle spectrum is given by \beq E^2_{k} =
(\varepsilon_k-\varepsilon_F)^2 + \Delta(k)^2, \eeq where
$\varepsilon_{k}=k^2/{2m^*}+U_0$ is the single-particle energy in
the effective mass approximation (EMA) and $U_0$ is mean field potential. 
$\varepsilon_F$ is the Fermi energy. In a consistent approach the gap equation has to
be coupled to the conservation of the particle number
 \beq \rho = 4 \sum_k \frac{1}{2} \big[ 1-\frac{\varepsilon_{k}
 -\varepsilon_F}{E_{k}}
\big], \eeq The pairing force, in principle, contains all
irreducible interaction diagrams, but here only the NN bare
interaction and the medium polarization insertions will be considered, as displayed in
Fig.~1. The bare two-particle interaction is
\beq V^{jst}(\kv,\kv') = N_0^{-1}\sum Y^*_{lm}(\hat k)
Y_{l'm'}(\hat k')C(lm,ss_z|jj_z)
C(l'm',ss'_z|jj_z)V_{ll'}^{jst}(k,k'), \eeq
where j, s and t are total angular momentum, spin and isospin.
\begin{figure}[tbh]
\centering
\includegraphics[angle=0,scale=0.2]{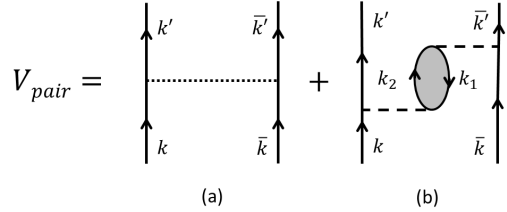}
\caption{Pairing interaction with screening: the first term in the
r.h.s. is the bare interaction, the second one is the induced
interaction, where the dashed bubble insertion is the series of
ring diagrams.} \label{mmm}
\end{figure}


\subsection{Induced interaction in RPA}
In this section, we discuss  the derivation of the vertex
corrections to the pairing interaction from the BS equation in the
RPA limit \cite{Dick83}
 \beq
 \mathcal{F}^T_{SM,SM'}(k,k';q) = G^{T}_{SM,SM'}(k,k';q)
+ \sum_{M''} \int \frac{d^4k''}{(2\pi)^4} \times
G^{T}_{SM,SM''}(k,k'';q)
\lambda(k'',q)\mathcal{F}^{T}_{SM'',SM'}(k'',k';q), \eeq
 where the $k$, $k'$ and $q$ stand for energy-momentum and energy-momentum
transfer, respectively, and $S$ and $T$ are total p-h spin (with
z-projection M) and isospin, respectively. $\lambda(k,q)$ is the
free polarization propagator \cite{Fetter}. The solution of the BS
equation assumes an algebraic form and can be solved analytically
\cite{friman} in the Landau limit, where energy and momentum lie
on the Fermi surface and energy-momentum transfer $q=0$. In that
limit the leading term $G(\bm k,\bm k';0)$ (spin-isospin here
omitted for simplicity) depends only on the angle $\theta$ between
$\bm{k}$ and $\bm{k'}$, expressed in terms of Landau-Migdal
parameters (expanded in partial waves), and can be written \bea
 G(\bm{k},\bm{k'})= N^{-1}_0 \sum_l (F_l
 + F'_l \bm\tau_1 \cdot \bm\tau_2
 + G_l \bm\sigma_1 \cdot \bm\sigma_2
 + G'_l \bm \sigma_1 \cdot \bm\sigma_2 \bm\tau_1 \cdot \bm\tau_2
 +\frac{q^2}{k^2_F}H_l S_{12}(q)+
 \frac{q^2}{k^2_F}H'_l S_{12}(q)\bm\tau_1 \cdot\bm\tau_2)
  P_l(cos\theta)
 \eea
where $2\bm{q} = \bm{k} -\bm{k'}$ is the relative momentum and $
 S_{12}$ the tensor operator, $S_{12}(q) = 3(\vec S\cdot\hat{q})^2 - S^2$. 
The $P_l(cos\theta)$ are the Legendre polynomials.
The inclusion of the tensor Landau-Migdal parameters is motivated by the fact
that the interaction contains a strong tensor component in the
$^3SD_1$ channel, as already pointed out in the Introduction.

In the BS equation the choice of the driving term  plays a crucial
role. In principle it contains all irreducible processes of the
interaction. The simplest approximation is to take the bare
interaction itself, but, to prevent the divergences related to the
hard-core of the nuclear force, we adopted the Brueckner
$G$-matrix calculated in the Brueckner-Hartree-Fock (BHF)
approximation. The relation between the $G$-matrix and the
Landau-Migdal parameters is presented in the Appendix.

In order to derive the BS equation in the Landau limit, we follow
closely Ref.~\cite{friman}. After expanding in partial waves the
p-h interaction $\mathcal{F}$ (the same for the leading term)
\begin{align}
\mathcal{F}^T_{SM_S,SM'_S}(\kv,\kv';0)= N_0^{-1}
\sum_{\substack{lml'm'\\JM}}\frac{4\pi Y^*_{lm}(\hat
k)Y_{l'm'}(\hat k')}{[(2l+1)(2l'+1)]^{1/2}}
 \langle lm\ SM_S|JM\rangle
\langle l'm'\ SM'_S|JM\rangle \mathcal{F}^{SJT}_{ll'},
\end{align}
the BS equation goes over into the  algebraic equation for the
$\mathcal{F}^{SJT}_{ll'}$ matrix elements
 \beq \mathcal{F}^{SJT}_{ll'} =
G^{SJT}_{ll'} -
\sum_{l''}\frac{1}{2l''+1}G^{SJT}_{ll''}\mathcal{F}^{SJT}_{l''l'}\cdot
\eeq\
where $J$ is the total angular momentum. The matrix elements
$G^{SJT}_{ll''}$ are the coefficients of the partial-wave expansion
of the leading term. Its expression in terms  of the Landau-Migdal
parameters is reported in the Appendix.
 In the case of $S=0$, all partial-wave matrix elements are
diagonal, because the tensor force does not affect the scalar
Landau-Migdal parameters and we simply get the well known
expression \beq
\mathcal{F}^{0JT}_{ll} =
\frac{G^{0JT}_{ll}}{1+G^{0JT}_{ll}/(2l+1)} \,,
\eeq where $G^{0J0}_{ll}=F_l$, $G^{0J1}_{ll}=F'_l$ and $J=l$.

In the case of $S=1$, off-diagonal matrix elements also exist due
to the coupling between vector and tensor Landau-Migdal parameters
as shown in the Appendix . But only two different angular momenta
$(l,l+2)$ can couple because we have to couple $l$ and $S$ to good $J$.  
The explicit expression of the matrix
elements $\mathcal{F}^{1JT}_{ll}$ (for $l=J\pm1$ and $l'=J\mp1$) of the effective
interaction is 
\bea
\mathcal{F}^{1JT}_{ll}  &=&
D^{-1}\big[G^{1JT}_{ll}\big(1+\frac{G^{1JT}_{l'l'}}{2l'+1}\big)
                -\frac{(G^{1JT}_{ll'})^2}{2l'+1}\big], \nonumber\\
\mathcal{F}^{1JT}_{ll'} &=& D^{-1}G^{1JT}_{ll'}\,,
\eea where
\beq D = \big(1+\frac{G^{1JT}_{ll}}{2l+1}\big)
\big(1+\frac{G^{1JT}_{l'l'}}{2l'+1}\big)
   - \frac{(G^{1JT}_{ll'})^2}{(2l+1)(2l'+1)}
\eeq Notice again that only two different angular momenta at most can
couple together, i.e. $l=l'$ or $|l-l'|=2$.

\subsection{Induced interaction}

For application to the gap equation the particle-hole (p-h) interaction must be
converted into particle-particle (p-p) interaction and then the induced part must be
taken out (second diagram on the r.h.s. of Fig.~1).  
The spin-isospin transformation is given by
\begin{align}
\mathcal{F}_{pp}^{st}(\bm{q},\bm{P}) = (-)^{1+t}\sum_{ST} (2T+1)
\left\{
\begin{array}{ccc}
\frac{1}{2} & \frac{1}{2} & T \\
\frac{1}{2} & \frac{1}{2} & t
\end{array} \right\}\sum_{MM'mm'} \{ SM,SM'|sm,sm'\} \mathcal{F}^T_{SM_S,SM'_S}(\kv,\kv';0),
\end{align}
where $\{SM,SM'|sm,sm'\}$ is the spin transformation bracket and
$\bm{P}$ is the total momentum \cite{Dick81}. For application to
the np pairing interaction in the $^3SD_1$ channel the
$\mathcal{F}^{01}_{ll'}$ partial waves with $l,l'=0,2$ have to be
projected out from the expansion of $\mathcal{F}^{st}(\bm{q},\bm{
P})$.


\section{Numerical results}

The numerical evaluation  of the medium polarization effects
starts from the $G$-matrix calculated in the BHF approximation
with the Argonne AV18 as two-body interaction and the consistent
meson-exchange three body force \cite{3bf}.

From the $G$-matrix expansion of the self-energy  the dispersion
effects of the mean-field are included in the effective mass
approximation (EMA) and the depletion of the Fermi surface is also
approximated by the Z-factors \cite{Dong13}.

The medium polarization is described by the BS equation, solved in the
RPA, where the  $G^{ph}$-matrix is the input, so that p-p short-range
correlations and p-h long-range collective excitations of the nuclear matter
are simultaneously treated in a unified manner.

Finally the p-p interaction induced by the medium polarization
is added to the bare interaction and the gap equation is solved.

\begin{figure}[tbh]
\centering
\includegraphics[angle=0,scale=0.25]{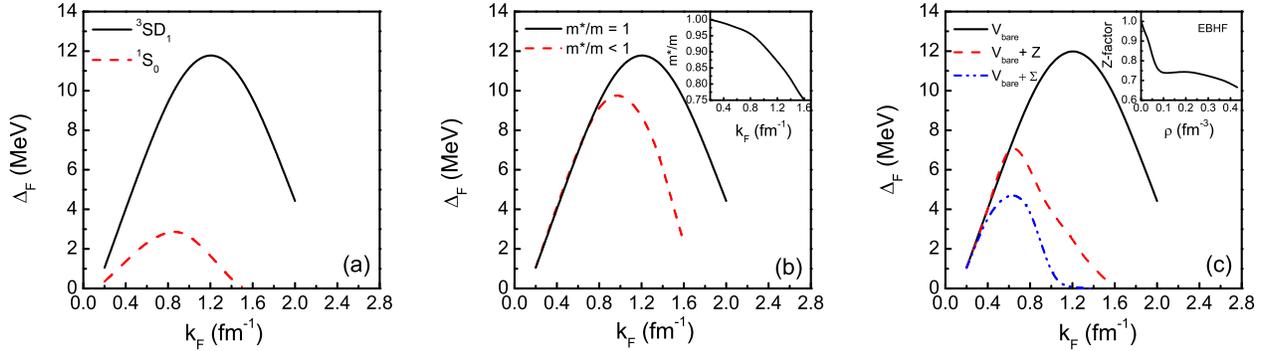}
\caption{Energy gap with self-energy effects. Left: Comparison between $^3SD_1$ and $^1S_0$ gaps
from bare interaction. Middle: $^3SD_1$ gaps with single particle
spectrum in the EMA (effective mass vs density in the
inset). Right: Energy gap with depleted Fermi surface (Z-factor vs
density in the inset).} \label{mmm}
\end{figure}

\subsection{Self-energy corrections}

As shown in Fig.~2a, the energy gap with only bare interaction
gives for spin-triplet $^3SD_1$ pairing a peak value the order of
$12$ MeV, which should be compared with the value of $3$ MeV for
spin-singlet $^1S_0$ pairing \cite{lombardo}. The large difference
between the two gaps is justified by the exponential dependence on
the interaction strength of the solution of the BCS gap
equation \cite{Fetter}. In fig.~2b, the mean field dispersive effect
is also reported for comparison using the effective mass (see
inset) in the quasi-particle spectrum, according to Eq.(3). This
effect is well known \cite{cugnon}: the gap magnitude gets reduced
and pairing density range is also shifted towards low densities,
where $ m^*/m\approx 1$. Additional reduction of the gap is
obtained when including the depletion of the Fermi sphere, as
shown in Fig.~2c. The depletion is introduced via the
Z-factor \cite{Zdick,Dong13}, plotted in the inset of the figure.
This quenching effect is more pronounced since pairing strength is
exponentially dependent on $Z^2$. The two combined effects give
rise to a remarkable quenching of the gap in a density range
making the $^3SD_1$ pairing a surface effect like the $^1S_0$ one.
However the peak value of $^3SD_1$ energy gap is still over two
times larger than the $^1S_0$ with the same self-energy
approximation.
\begin{figure}
\centering
\includegraphics[angle=0,scale=0.25]{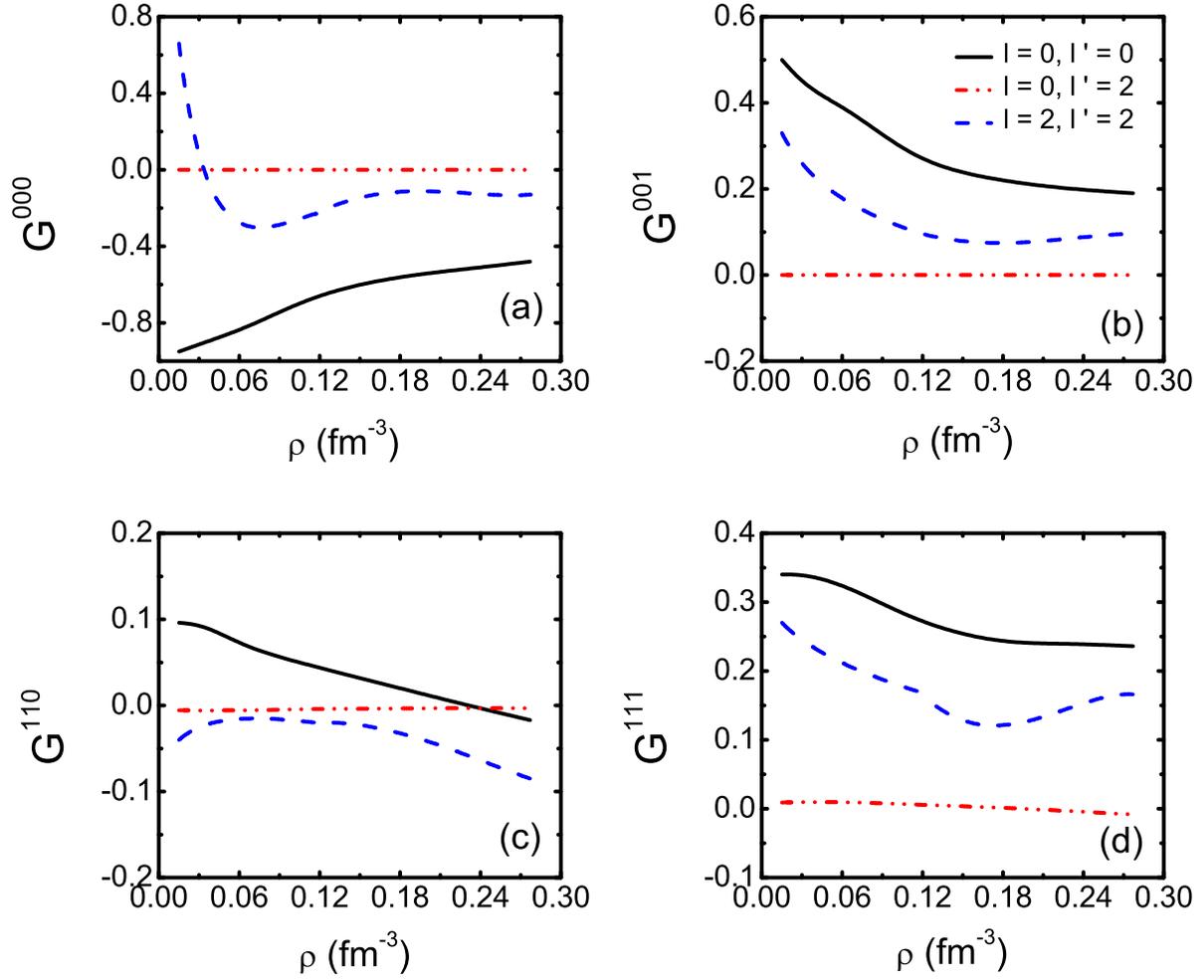}
\caption{Leading p-h interaction $\mathcal{F}^{SJT}_{ll'}$ from
G-matrix in SD channel.} \label{mmm}
\end{figure}
\begin{table}
\renewcommand{\arraystretch}{1.5}
\begin{ruledtabular}
\vspace{1mm}
\begin{tabular}{ccccc}
$\rho$  &$k_F$ &${SS}$  &${DD}$    &${SD}$ \\
$(fm^{-3})$ &$(fm^{-1})$ &$(MeV\cdot fm^{3})$ &$(MeV\cdot fm^{3})$ &$(MeV\cdot fm^{3})$\\
\hline
0.277 &1.60 &  -0.70  &-0.03 &  0.04  \\
0.228 &1.50 &  -1.14  &-0.04 &  0.02  \\
0.186 &1.40 &  -1.63  &-0.04 &  0.00  \\
0.175 &1.36 &  -1.65  & 0.05 & -0.01  \\
0.117 &1.20 &  -3.43  &-0.05 & -0.04  \\
0.068 &1.00 & -13.03  &-0.07 & -0.08  \\
0.035 &0.80 & -22.32  & 0.00 & -0.10  \\
\end{tabular}
\end{ruledtabular}
\caption{p-p induced interaction $(\mathcal{F}_{pp}^{10})_{ll'}$
in the $^3SD_1$ channel.}
\end{table}

\subsection{Induced interaction}

The p-p matrix elements of the $G$-matrix in the SD channel are
calculated from the BHF approximation with the same two and three
body force like the self-energy. The p-p matrix elements are
transformed into p-h matrix elements, expressed in terms of
Landau-Migdal parameters as shown in the Appendix. For such a
purpose the Landau limit has been adopted, where the
energy-momentum transfer is assumed to be vanishing. Since the SD
components of the $G$-matrix derive from the tensor part of the bare
interaction the additional H and H' Landau-Migdal parameters have
been introduced in the p-h effective interaction. In Fig.~3 the SD
partial-wave of the BHF Landau-Migdal parameters are plotted as a
function of the density, Eqs.(A6-A10). The zero-order diagonal
components are the Landau-Migdal parameters from the BHF
$G$-matrix with no tensor force effect. The $S=1$ partial-wave
components are affected by the tensor Landau-Migdal parameters,
but their effect is small. It follows that the off-diagonal matrix
elements are even smaller. The main contribution comes from  the
isoscalar and isovector density fluctuations ($S=0$) as expected.

From the solution of the BS equation in the Landau limit with BHF
Landau-Migdal parameters shown in Fig.~3 as input, the effective
interaction is determined and transformed in the p-p
representation, according to Eq.(13). The matrix elements of the
$^3SD_1$ induced part (second diagram of the r.h.s. of Fig.~4) are
reported in Table I. It easily  seen that the dominant
contribution is concentrated in the $S=0$ isoscalar matrix
element. This contribution is attractive much the same as for the
spin-singlet pairing in symmetric nuclear matter \cite{cao06}.

\begin{figure}
\centering
\includegraphics[angle=0,scale=0.15]{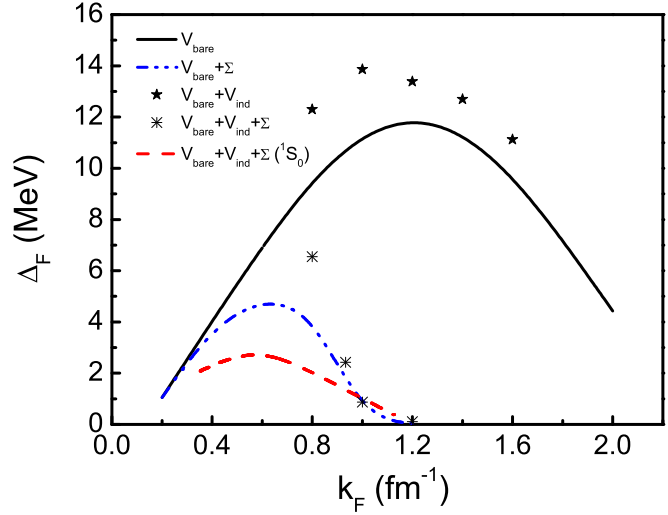}
\caption{Comparison among $^3SD_1$ gaps from RPA induced
interaction and previous effects
and $^1S_0$ in full calculations.} \label{mmm}
\end{figure}

\subsection{Pairing gap from vertex corrections}

The p-p effective interaction is added to the pairing interaction
and the BCS equation is solved. The resulting gap vs. Fermi
momentum is displayed in Fig.~4 in comparison with the preceding
results. Two series of calculations have been performed: the first
one (upper stars) shows the effects on induced interaction without
self-energy corrections, the second one (lowest stars) is full
calculations, self-energy plus induced interaction. There was a
limit to the lower densities imposed by the missing convergence of
the BHF calculation of $G$-matrix. This is due to the singularity of
$G$-matrix in the density domain where large pair correlations are
expected to occur. This drawback requires a self-consistent
calculation of BCS equation and BHF calculation with
quasi-particle energy spectrum, that is beyond the scope of the
present investigation. The main result is that, due to the
attractive nature of the new term, the self-energy quenching is
reduced, but less than in the case of spin-singlet pairing. A
second main result is that the shift of the peak value at low
density produced by the self-energy is not affected by the induced
interaction, suggesting that  pairing is a surface phenomenon in
finite nuclei. Finally, it is worth of noticing that the
spin-triplet pairing in $^3SD_1$ channel is still much larger than
the spin-singlet pairing in $^1S_0$ channel, as clearly shown in
Fig.~4.

\subsection{Average pairing in nuclei from LDA}

To make contact with pairing in nuclei we have estimated the average gap in N=Z
nuclei from the Thomas-Fermi density corresponding to the states around the chemical potential $\mu$  defined
as follows \cite{schuck}
\bea
<\Delta(\mu)> = \int {d\vec r} \sum_i \frac{1}{g(\mu)}
\delta(\mu-\varepsilon_i) |\phi_i(\vec r)|^2 \Delta(r)
\eea
where $\phi_i(\vec r)$ is the single-particle wave function with energy
eigenvalue $\varepsilon_i$, $\Delta (r)$ is the nuclear matter gap
for the density $\rho(r)$ according to the local density approximation (LDA) and $g(\mu)$ is the level density at $\mu$.
It easy to show that, in the $\hbar\Rightarrow 0$ semiclassical
limit \cite{schuck}
\bea
<\Delta(\mu)> = \frac{\int{d^3\vec r}\Delta(r) \rho^{1/3}(\vec r)}
{\int{d^3\vec r} \rho^{1/3}(\vec r)}
\eea
where we take for the density the phenomenological one parametrized in \cite{shlomo}.
In Table II the results are reported for some N=Z nuclei. We see that screening substantially reduces the gap values which, however, remain stronger than in the $T=1$ channel. A word of caution is, however, in order: in finite nuclei the effect of collective surface modes may be quite different from what can be simulated with LDA. So the results from the latter should be taken only as a qualitative indication.

\begin{table}
\renewcommand{\arraystretch}{1.5}
\begin{ruledtabular}
\vspace{1mm}
\begin{tabular}{cccc}
 $A$  & $R(fm)$&$\Delta_0(MeV)$&$ \Delta (MeV)$\\
\hline
 40 & 3.83  &6.82 & 3.54    \\
100 & 5.20  &8.18 & 3.75    \\
200 & 6.50  &9.38 & 4.00    \\
\end{tabular}
\end{ruledtabular}
\caption{Average gaps $\Delta_0$ (no screening) and $\Delta$ (screening) in N=Z nuclei from LDA.
The density profiles are taken from Ref.~\cite{shlomo}.}
\end{table}

\section{Discussion and conclusions}

In this paper the spin-triplet $^3SD_1$ pairing in symmetric
nuclear matter has been discussed within the BCS theory with
medium polarization effects. On one hand, the self-energy
corrections reduce significantly the magnitude of the gap,
shifting the peak value to low density. On the other hand, the
induced interaction that is attractive almost in the full
asymmetry range, partially restores a higher magnitude of the gap
without additional squeezing of the density range of the
superfluid phase. The induced interaction has been calculated from
the RPA in the Landau limit, starting from the BHF p-h
interaction. In this fashion the long-range correlations are built
up on top of the short-range correlations from $G$-matrix. In this
approximation the main contribution comes from the scalar density
fluctuations, as expected. On the other hand, the feedback of the
vertex corrections on the other spin-isospin fluctuations can only
be treated within the framework of the induced interaction
approach \cite{babu} that is a task of further investigation.

The gaps obtained in the present approximation, as large as 2-3
times the magnitude of the spin-singlet pairing in the $^1S_0$
channel, provide a strong indication of the importance of the
medium polarization. The conclusion is that the $^3SD_1$
neutron-proton superfluid state in nuclear matter turns out to be
more stable than the $^1S_0$ neutron-neutron or proton-proton
superfluid state. This is in keeping with the calculations, where
it is found that N=Z heavy nuclei ($A>140$) np pair correlations
are stronger than nn or pp ones \cite{bertsch10}. Below this
threshold the pairing between like nucleons is found to be the
favored one, because the spin-orbit splitting in nuclei hinders np
pairing, but the present nuclear-matter calculations address the
problem whether the np pairing strength might be larger in the
spin triplet than singlet pairing state even for nuclei below
$A=140$. It would be a timely issue to study the competition between
spin-triplet pairing with medium screening effects and the
spin-orbit splitting in finite nuclei, considering that tools
already have been devised to face such a problem \cite{Idi}.

\begin{acknowledgments}
The authors are grateful to G. L. Col\`{o} and E. Vigezzi for
useful discussions. This work was supported by INFN post-doc
fellowship program and the National Natural Science Foundation of
China under Grant No.~11705109.
\end{acknowledgments}

\appendix
\section{Landau-Migdal parameters from BHF G-matrix}

The microscopic derivation of the Landau-Migdal parameters from
the BHF approximation is obtained converting the p-p $G$-matrix,
as calculated with the Brueckner-Bethe-Goldstone equation, into
the p-h representation. This procedure yields \cite{back,haen} \bea
(F, F') &=& \frac{1}{16} \sum_{st} (2t \pm 1)G^{st} \\(G, G') &=&
\frac{1}{16} \sum_{t} (2t \pm 1) (G^{1t}-G^{0t}) \\(H, H') &=&
\frac{1}{24}\frac{k_F^2}{q^2}\sum_{t} (2t \pm 1)(\tilde
G^{1t}_1-\tilde G^{1t}_0), \eea
where $G^{st}$ denotes the $G$-matrix with spin s and isospin t,
and $\tilde G^{st}_m$ is the same with $\hat q$ along the spin
quantization axis. The isoscalar (isovector) Landau-Migdal
parameters take the upper (lower) sign. Inverting the partial-wave
expansion of the leading term  we can determine the coefficients
\bea
 G^{SJT}_{ll'}&=& N_0 \frac{[(2l+1)(2l'+1)]^{1/2}}{4\pi}
\sum_{\substack{mm',M_sM'_s}} {[(2l+1)(2l'+1)]^{1/2}} \langle
lm\ SM_S|JM\rangle \langle l'm'\ SM'_S|JM\rangle \times \\
& &\int d{\hat k}d{\hat k'} Y_{lm}(\hat k) Y^*_{l'm'}(\hat k')
\langle SM_s,T|{G}^{ST}(\bm k,\bm k')| SM'_s,T\rangle,
\eea as a function of the Landau-Migdal parameters. For the $S=0$
component the calculation is straightforward, whereas for $S=1$ it
is quite tedious for the coupling between vector and tensor
Landau-Migdal parameters. It can be found  in the literature (see,
e.g. Refs.~\cite{nakayama,Dick83}. Below we report the matrix
elements needed for the calculation of vertex correction to the np
pairing interaction in the channel $^3SD_1$. For $T=0$ they are in
the order \bea
{G}^{000}_{00} &=& F_0\\
{G}^{020}_{22} &=& F_2  \\
{G}^{110}_{00} &=& G_0 \\
{G}^{110}_{22} &=& G_2 -
\frac{1}{4}(\frac{7}{3}H_1-2H_2+ \frac{3}{7}H_3)\\
{G}^{110}_{02} &=& - \frac{\sqrt{10}}{12}(3H_0-2H_1+
\frac{3}{5}H_2) \eea For $T=1$ the isoscalar Landau-Migdal
parameters must be replaced by the corresponding isovector ones,
$(F \rightarrow F',...)$.


\end{document}